# Experimental test of the third quantization of the electromagnetic field

J.D. Franson, Physics Dept., University of Maryland Baltimore County, Baltimore, MD 21250 U.S.A.

Each mode $j$ of the electromagnetic field is mathematically equivalent to a harmonic oscillator with a wave function $\psi_j(x_j)$ in the quadrature representation. An approach was recently introduced in which $\psi_j(x_j)$ was further quantized to produce a field operator $\hat{\psi}_j(x_j)$ [J.D. Franson, Phys. Rev. A **104**, 063702 (2021)]. This approach allows a generalization of quantum optics and quantum electrodynamics based on an unknown mixing angle $\gamma$ that is somewhat analogous to the Cabibbo angle or the Weinberg angle. The theory is equivalent to conventional quantum electrodynamics for $\gamma = 0$, while it predicts a new form of inelastic photon scattering for $\gamma \neq 0$. Here we report the results of an optical scattering experiment that set an upper bound of $|\gamma| \leq 1.93 \times 10^{-4}$ rad at the 99% confidence level, provided that the particles created by the operator $\hat{\psi}_j(x_j)$ have negligible mass. High-energy experiments would be required to test the theory if the mass of these particles is very large.

## I. INTRODUCTION

Each mode $j$ of the electromagnetic field is mathematically equivalent to a harmonic oscillator [1-5]. It is customary in quantum optics to introduce operators $\hat{x}_j$ and $\hat{p}_j$ that are proportional to the electric field of mode $j$ and its time rate of change. This allows the definition of a wave function $\psi_j(x_j)$ in the so-called quadrature representation of mode $j$ [1, 2, 6-8], which is used extensively to calculate the nonclassical properties of the field [9,10]. An approach in which the wave function $\psi_j(x_j)$ is further quantized to produce a field operator $\hat{\psi}_j(x_j)$ was recently proposed [11]. Since the electromagnetic field is already second-quantized, this corresponds to an additional or third quantization. This approach is required for a complete description of quantum optics in the Heisenberg picture [11].

The third quantization approach is equivalent to conventional quantum optics and quantum electrodynamics if we use the standard Hamiltonian. But an analogy with symmetry breaking in elementary particle theory suggests a generalization of quantum electrodynamics [11] based on a mixing angle $\gamma$ that is somewhat analogous to the Cabibbo angle [12] or the Weinberg angle [13]. The generalized theory predicts new effects that could be tested experimentally if $\gamma \neq 0$.

There are a number of recent elementary particle experiments whose results appear to be inconsistent with the predictions of the standard model, which has generated considerable interest in theories that go beyond the standard model (BSM) [14-18]. The third quantization of the electromagnetic field is an example of a BSM theory, even though it was originally developed for use in quantum optics. For example, the particles created by the field operator $\hat{\psi}_j(x_j)$ may be candidates for the dark matter inferred from astronomical observations.

Here we report the results of an optical experiment in which the generalized theory predicts a new form of inelastic photon scattering if $\gamma \neq 0$. The experimental results set an upper bound of $|\gamma| \leq 1.93 \times 10^{-4}$ rad at the 99% confidence level, provided that the hypothetical particles created by the field operator $\hat{\psi}_j^{\dagger}(x_j)$ have negligible mass $m$. High-energy experiments would be required to test the theory if the value of $m$ is very large.

The remainder of the paper is organized as follows. Section II provides a brief review of the third-quantization approach and the role of the mixing angle $\gamma$. The design of the experimental apparatus and the experimental approach are described in Section III. The experimental results and the bound on the value of $\gamma$ are discussed in Section IV. A summary and conclusions are presented in Section V, including a discussion of the need for high-energy experiments to investigate the possibility that these particles have a large mass.

## II. OVERVIEW OF THE THIRD QUANTIZATION APPROACH

Each mode $j$ of the electromagnetic field is mathematically equivalent to a harmonic oscillator [1-5]. We can think of each of these harmonic oscillators as containing a single hypothetical particle whose excited states $|n_j\rangle$ correspond to the presence of $n_j$ photons in mode $j$ of the field. This is illustrated in Fig. 1(a), along with the usual photon annihilation and creation operators $\hat{a}_j$ and $\hat{a}_j^\dagger$.

The third-quantization approach [11] introduces a new operator $\hat{c}_{jn}^\dagger$ that creates an additional hypothetical particle of that kind in state $|n_j\rangle$ of the harmonic oscillator representing mode $j$, as illustrated in Fig. 1(b). For lack of a better term, these hypothetical particles are referred to as oscillatons. The field operator $\hat{\psi}_j(x_j)$ for a single mode of the electromagnetic field can then be defined as

$$\hat{\psi}_j(x_j) \equiv \sum_n \hat{c}_{jn} \phi_n(x_j). \tag{1}$$

Here $\phi_n(x_j)$ is the $n$th energy eigenstate of the harmonic oscillator potential.

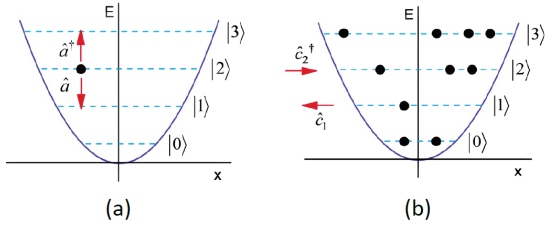

FIG. 1. (a) In conventional quantum optics, each mode of the electromagnetic field is equivalent to a harmonic oscillator that contains a single hypothetical particle (an oscillaton) represented by a black dot. The usual photon creation and annihilation operators $\hat{a}^\dagger$ and $\hat{a}$ increase or decrease the energy of the particle by $\hbar\omega$, which corresponds to the emission or absorption of a photon. (b) In the third-quantized theory, the operator $\hat{c}_n^\dagger$ creates an additional hypothetical particle of that kind in state $|n\rangle$. (From Ref. [11]).

New photon annihilation and creation operators $\hat{a}'_j$ and $\hat{a}'^\dagger_j$ can be defined as

$$\begin{aligned}\hat{a}'_j &\equiv \sum_{n=1}^\infty \sqrt{n}\, \hat{c}_{j,n-1}^\dagger \hat{c}_{jn}, \\ \hat{a}'^\dagger_j &\equiv \sum_{n=0}^\infty \sqrt{n+1}\, \hat{c}_{j,n+1}^\dagger \hat{c}_{jn}.\end{aligned} \tag{2}$$

These operators are equivalent to the usual photon creation and annihilation operators if there is a single oscillaton as in Fig. 1(a).

The vector potential operator $\hat{\mathbf{A}}(\mathbf{r})$ can then be defined as usual by [19]

$$\hat{\mathbf{A}}(\mathbf{r}) = \sum_{j,\varepsilon_j} \sqrt{\frac{2\pi\hbar c^2}{\omega_j L^3}} \left(\boldsymbol{\varepsilon}_j \hat{a}'_j e^{i\mathbf{k}_j\cdot\mathbf{r}} + \boldsymbol{\varepsilon}_j^* \hat{a}'^\dagger_j e^{-i\mathbf{k}_j\cdot\mathbf{r}}\right). \tag{3}$$

Here $\boldsymbol{\varepsilon}_j$ are two orthogonal polarization vectors, $L$ is the length used for periodic boundary conditions, and $c$ is the speed of light.

The standard Hamiltonian for the interaction of charged particles with the electromagnetic field can be written in the Coulomb gauge in the form [20]

$$\hat{H}' = -\frac{1}{c}\int d^3\mathbf{r}\, \hat{\mathbf{j}}(\mathbf{r})\cdot \hat{\mathbf{A}}(\mathbf{r}), \tag{4}$$

where $\hat{\mathbf{j}}(\mathbf{r})$ is the current carried by an electron or other particle. This Hamiltonian conserves the number of oscillatons and gives results that are equivalent to conventional quantum optics and quantum electrodynamics.

More general Hamiltonians need not conserve the number of oscillatons. An example of such a theory was suggested in Ref. [11], where it was shown that an interaction between the oscillatons and another boson B with a large mass would produce a Bogoliubov transformation [21, 22] of the form

$$\begin{aligned}\hat{c}_{jn} &\to \hat{c}'_{jn} = \beta\left(\cos\gamma\, \hat{c}_{jn} + \sin\gamma\, \hat{c}_{jn}^\dagger\right) \\ \hat{c}_{jn}^\dagger &\to \hat{c}'^\dagger_{jn} = \beta\left(\sin\gamma\, \hat{c}_{jn} + \cos\gamma\, \hat{c}_{jn}^\dagger\right).\end{aligned} \tag{5}$$

The constant $\beta = 1/(\cos^2\gamma - \sin^2\gamma)^{1/2}$ maintains the commutation relations, while the mixing angle $\gamma$ is somewhat analogous to the Cabibbo angle [12] or the Weinberg angle [13]. As a result of the interaction, the bare oscillatons no longer correspond to the true eigenstates of the system.

Although the assumed interaction with a massive boson may seem speculative, it is closely analogous to the symmetry breaking mechanisms that occur in the standard model of elementary particle theory. For example, the three-dimensional coupling matrix described in the appendix of Ref. [11] is analogous to the Cabibbo-Kobayashi-Maskawa matrix for quarks or the Maki-Nakagawa-Sakata matrix for neutrinos.

It can be shown [11] that the interaction Hamiltonian of Eq. (4) combined with the Bogoliubov transformation of Eq. (5) can create or annihilate a pair of oscillatons while emitting or absorbing a photon. This process can be understood using second-order perturbation theory [11], where there are two interactions with the massive boson field. The first interaction creates a new oscillaton and a virtual B particle, while the second interaction annihilates the B particle and creates a second oscillaton, resulting in the creation of a pair of oscillatons. The Bogoliubov transformation describes this in a more formal way.

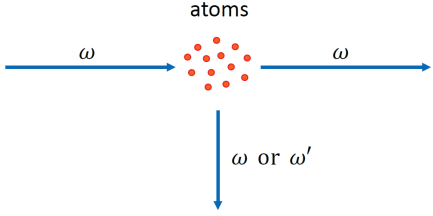

FIG. 2. Inelastic photon scattering predicted by the generalized theory of Eq. (5). Photons at frequency $\omega$ are incident on a cloud of two-level atoms. Some of the photons are scattered through a 90° angle with final frequencies of $\omega$ or $\omega' = \omega/2$. Energy is conserved in the latter case by the creation of a pair of oscillatons. The predicted ratio of the two scattering rates can be used to set an upper bound on the mixing angle $\gamma$. (From Ref. [11].)

This prediction can be tested using the photon scattering experiment outlined in Fig. 2. Here an incident photon at frequency $\omega$ is scattered by a two-level atom to produce a single photon at a frequency of $\omega$ or $\omega' = \omega/2$. Energy is conserved in the latter case by the creation of a pair of oscillatons, where we have assumed for the time being that the mass of the oscillaton is negligibly small.

The ratio $R$ of the inelastic scattering rate at frequency $\omega'$ divided by the usual elastic scattering rate at frequency $\omega$ is predicted by the generalized theory to be given by $R = 4\gamma^2$ [11]. Thus an experiment of this kind can set an upper bound on the value of the mixing angle $\gamma$.

### III. EXPERIMENTAL APPARATUS

The experimental apparatus used to test the generalized theory is outlined in Fig. 3. Three fiber-coupled diode lasers with wavelengths of 635, 780, and 1550 nm could be connected one at a time to a fiber port in front of lens $L_1$, which was mounted in a three-axis micropositioner. A collimated laser beam with a diameter of 1 mm could be produced by adjusting the position of $L_1$. Two pinholes $P_1$ and $P_2$ defined the central optical axis of the experiment, and mirrors $M_1$ and $M_2$ could be used to center the laser beams through the two pinholes. This ensured that the same optical path would be followed regardless of which laser was connected to the collimator.

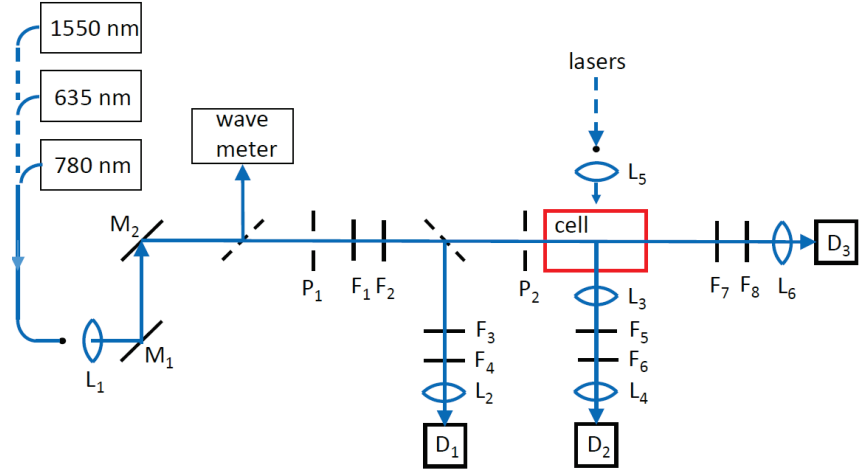

FIG. 3. Experimental apparatus used to measure the ratio of the inelastic scattering rate at 1560 nm divided by the usual elastic scattering rate at 780 nm. One of three diode lasers could be coupled into the experiment using an optical fiber. Two of the lasers were only used for initial alignment, while a frequency-stabilized laser at 780 nm was used in the actual scattering measurements. Detector $D_1$ was used to correct for variations in the laser power, detector $D_2$ measured the scattering rates, and detector $D_3$ measured the intensity of the light transmitted through a rubidium cell. $M_1$ and $M_2$ were alignment mirrors, $P_1$ and $P_2$ were pinholes, $F_1$ through $F_8$ were filters, and $L_1$ through $L_6$ were lenses, as described in the text.

The visible laser at 635 nm and the 1550 nm laser were used only for initial alignment and testing, as will be described below. The 780 nm laser was a frequency-stabilized (New Focus Velocity) diode laser whose frequency could be scanned over one of the hyperfine absorption lines of single-isotope $^{87}$Rb atoms contained in a fused silica cell. An oven was used to heat the rubidium cell over a range of temperatures up to 50° C.

A glass plate was used to couple a small fraction of the laser power into a precision wave meter (HighFinesse model WS-U) that could measure the frequency of the laser light with a precision of approximately 10 Mz. Filter $F_1$ was a variable neutral density filter that could be used to adjust the laser power to the desired value. Filter $F_2$ consisted of a pair of narrow-band interference filers with central wavelengths of 780 nm and bandwidths of 5 nm. Filter $F_2$ and the remainder of the apparatus were enclosed in a light-tight box, which prevented any significant amount of light from reaching the detectors from the outside environment.

Another glass plate coupled a small fraction of the laser power into detector $D_1$, which was a silicon photodiode used to measure the power from the laser. This allowed the scattering rates to be corrected for small variations in the laser power. Filter $F_3$ was a neutral density filter used to prevent detector $D_1$ from saturating, while filter $F_4$ was a narrow-band interference filter centered at 780 nm. An additional pair of mirrors (not shown in the figure) was used to center the laser beam onto detector $D_1$.

Detector $D_2$ was a low-noise indium gallium arsenide photodetector (Thorlabs model PDF 10C) that was used to measure the intensity of the scattered light. This detector had an internal preamplifier and a noise equivalent power of $7 \times 10^{-15}$ W$/\sqrt{\text{Hz}}$, which allowed power levels as low as $1 \times 10^{-16}$ W to be measured using time-averaging techniques described below. This detector was used to measure the usual elastic scattering at 780 nm as well as the inelastic scattering at 1560 nm, even though the quantum efficiency was much less at 780 than 1560 nm. This difference in quantum efficiency was taken into account when calculating the ratio of the two scattering rates.

Several single-photon avalanche diode (SPAD) detectors suitable for use at 1560 nm were also available in the laboratory, but their quantum efficiency was lower than that of detector $D_2$ and their active area was two orders of magnitude smaller. As a result, $D_2$ was the best choice for this application.

Lens $L_3$ formed a collimated beam from the light scattered by the rubidium atoms, while lens $L_4$ focused the beam onto detector $D_2$. When measuring the elastic scattering rate at 780 nm, filter $F_5$ was a neutral density filter that reduced the intensity of the scattered light to avoid saturating detector $D_2$, while filter $F_6$ was a narrow-band interference filter with a central wavelength of 780 nm.

When measuring the inelastic scattering rate at 1560 nm, $F_5$ was replaced with a set of three long-pass filters that attenuated the scattered light at 780 nm by approximately 12 orders of magnitude. Filter $F_6$ was also replaced with a narrow-band interference filter with a central wavelength of 1560 nm and a bandwidth of 12 nm. A sequence of narrow-band filters that covered the range from 1530 through 1580 nm could also be used to measure the wavelength dependence of any scattered light.

Lenses $L_3$ and $L_4$ were achromatic doublets optimized for wavelengths near 780 nm during the measurement of the elastic scattering rate. These were replaced with lenses optimized for wavelengths near 1560 nm during the measurement of the inelastic scattering rate. The differences in the loss or attenuation of all of these elements was taken into account when calculating the ratio of the two scattering rates.

Detector $D_3$ was a silicon photodiode that measured the intensity of the light transmitted through the rubidium cell as the laser frequency was continuously scanned over a 5 GHz range centered on one of the hyperfine absorption lines. This was used to stabilize the frequency of the laser, which would otherwise drift by approximately 0.1 GHz/h due to changes in the room temperature. Filters $F_7$ and $F_8$ were neutral density and narrow-band interference filters used to limit the light entering detector $D_3$. Another pair of mirrors (not shown) was used to center the beam onto detector $D_3$.

The use of narrow-band interference filter $F_6$ required that the light travelling toward detector $D_2$ be well collimated. This was achieved by initially removing the rubidium cell, lenses $L_3$ and $L_4$, filters $F_5$ and $F_6$, and detector $D_2$. The visible 635 nm diode laser was then connected to a fiber port behind lens $L_5$, which was mounted on a three-axis micropositioner and used to produce a collimated beam. Lens $L_4$ was then put into position and detector $D_2$ was mounted on a three-axis micropositioner, which allowed it be placed at the

5focal point of fixed lens $L_4$. Detector $D_2$ had a small active area (a diameter of 0.5 mm) and was therefore most sensitive to light that was collimated in the direction determined by lens $L_5$.

The 635 nm laser was then connected to the fiber port behind lens $L_1$ and a diffuse reflector was placed in the laser beam where the cell would normally be located. This produced a wide angle of scattered light that simulated the subsequent scattering by the rubidium atoms. Lens $L_3$ was then mounted on a three-axis micropositioner and adjusted to focus the scattered light onto detector $D_2$. The oven and rubidium cell were then put back into position. This procedure ensured that the light scattered by the rubidium atoms would be properly collimated with the optimal detection efficiency. The entire process could be repeated using the 780 and 1550 nm lasers instead.

Perhaps the biggest concern in the design of the experiment was the possibility that scattered 780 nm photons could produce fluorescence in the walls of the cell, lens $L_3$, or filter $F_5$. Some of the fluorescence might pass through the subsequent long-pass and narrow-band filters, giving a spurious signal. Fluorescence of this kind would be expected to have a large bandwidth and could be distinguished in that way from any inelastic scattering at 1560 nm. No fluorescence of that kind was detected.

### IV. EXPERIMENTAL RESULTS

The frequency of the 780 nm diode laser could be controlled by applying a voltage to a piezoelectric element in the laser head. The output of a signal generator was used to modulate the laser frequency through a range of 5 GHz centered on one of the hyperfine transitions with a period of 10 s. The outputs of all three detectors along with the voltage from the frequency generator were input to a digital storage oscilloscope. The output of detector $D_2$ was amplified by a gain of 10 or 100 before being input to the oscilloscope, depending on the measurement being made. After digitizing the signals, the storage oscilloscope averaged them for 30 min and the results were then transferred to a computer for analysis and plotting. The results from 9 data runs of that kind were averaged to further reduce the statistical uncertainties.

Fig. 4 shows the transmitted intensity at 780 nm as measured by detector $D_3$, plotted as a function of the detuning of the laser from the resonant frequency of the rubidium atoms. The oven containing the rubidium cell was maintained at a temperature of 37 C°, which was the case for all of the subsequent measurements. Here the laser power incident on the rubidium cell had been attenuated to a relatively low level of 0.1 $\mu$W, which is below the atomic saturation intensity. It can be seen that a substantial fraction of the incident laser power was scattered out of the original beam under these conditions.

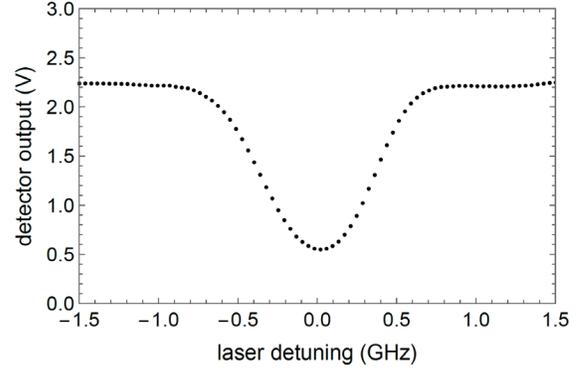

Fig. 4. Transmission through the rubidium cell as measured by the output of detector $D_3$ (in volts) for a relatively small laser power of 0.1 $\mu$W. The transmission is plotted as a function of the detuning of the 780 nm laser from the rubidium resonance frequency.

The elastic scattering rate at 780 nm increases as the laser power incident on the rubidium cell is increased, although the increase is not linear due to saturation and power broadening at higher intensities. Fig. 5 shows the transmission through the rubidium cell when the maximum available laser power (10 mW) was incident on the cell. It can be seen that a smaller fraction of the incident power was scattered, but the total rate of photon scattering was still much higher than at lower power levels. This laser intensity was used for all of the subsequent measurements.

Fig. 6 shows the output of detector $D_2$ for elastic scattering at 780 nm, plotted as a function of the laser detuning. Here filter $F_5$ was used to attenuate the signal by a factor of 620 in order to avoid saturating the detector, since the scattering at 780 nm was relatively strong. Filter $F_6$ was a narrow-band filter centered on 780 nm. The signal from detector $D_2$ was amplified by an external gain of 10 in this case. It can be seen that the usual elastic scattering at 780 nm can be readily observed even after being strongly attenuated.

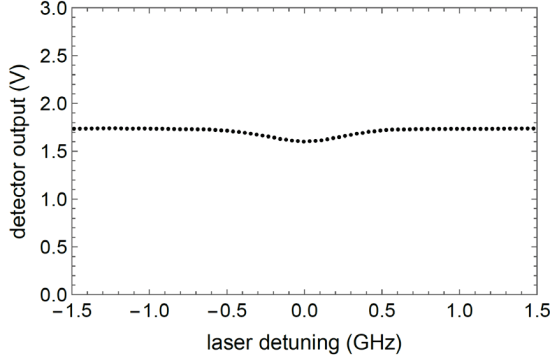

Fig. 5. Transmission through the rubidium cell as measured by the output of detector $D_3$ (in volts) for a laser power of 10 mW incident on the rubidium cell. Neutral density filter $F_7$ was used to attenuate the signal in order to avoid saturation of the detector. Although a smaller fraction of the incident power was scattered than in Fig. 4, the total rate of photon scattering was much higher.

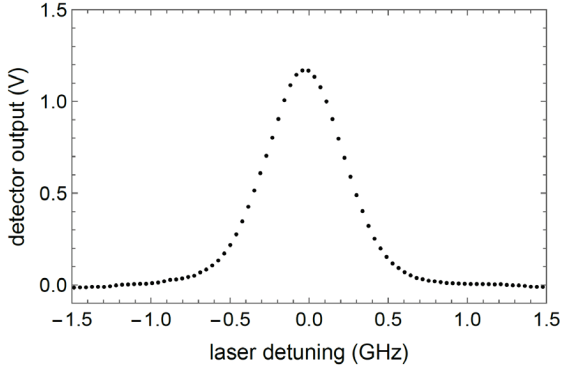

Fig. 6. Elastic scattering from the rubidium cell at a wave length of 780 nm as measured by the output of detector $D_2$ (in volts). The laser frequency was scanned over a range of 5 GHz centered on one of the rubidium hyperfine lines. The signal was attenuated by a factor of 620 using filter $F_5$ in order to avoid saturating the detector.

Fig. 7 shows the output of detector $D_2$ when the apparatus was configured to measure inelastic scattering at 1560 nm. Here the incident photons had a wavelength of 780 nm as before, but filter $F_5$ was replaced by a set of long-pass filters that strongly absorbed the scattered 780 nm photons. Filter $F_6$ was also replaced by a narrow-band filter centered on a wavelength of 1560 nm with a bandwidth of 12 nm. In contrast with the data of Fig 6, no neutral density filters were used and the output of detector $D_2$ was amplified by a factor of 100 instead of 10. In addition, the quantum efficiency of the detector was a factor of 5 higher at 1560 nm than it was for 780 nm. Nevertheless, it can be seen that there is no apparent peak in the data corresponding to inelastic scattering at 1560 nm.

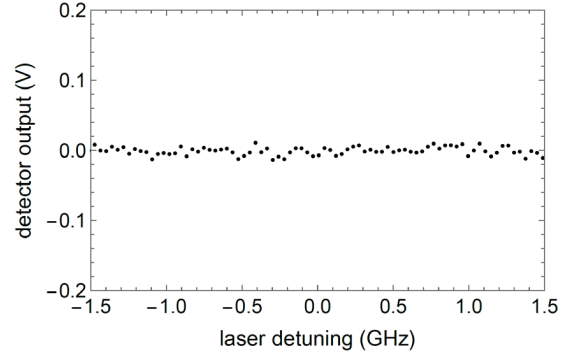

Fig. 7. Inelastic scattering results at a wavelength of 1560 nm as measured by the output of detector $D_2$ (in volts). Here filter $F_6$ had a narrow bandwidth at a central wavelength of 1560 nm. These measurements did not include the attenuation used in the 780 nm measurements of Fig 6, and the quantum efficiency and amplifier gain were also much larger than was the case for the 780 nm measurements. No significant signal was observed at 1560 nm.

The 780 nm scattering data of Fig. 6 were fit to a Gaussian with an adjustable peak $V_{780}$ in the detector output voltage and an adjustable width. The 1560 nm data of Fig. 7 were also fit to a Gaussian using an adjustable peak and the same width as was obtained from the fit to Fig. 6. The estimated value of the inelastic scattering "peak" was actually negative and consistent with zero, given the noise in the detector.

As a result, the upper bound on the scattering ratio $R$ was based on the uncertainty in the fit to the inelastic scattering data. The ratio was adjusted to take into account the attenuation from the neutral density filter in the 780 nm measurements, as well as the differences in the transmission losses of the various optical elements, the quantum efficiencies, and the amplifier gains. At the 99% confidence level (three standard deviations), the experimental bound on the value of $R$ is given by

$$R \leq 3 \frac{V_{1560}}{V_{780}} \frac{A_{1560}}{A_{780}} \frac{L_{1560}}{L_{780}} \frac{\eta_{780}}{\eta_{1560}} \frac{g_{780}}{g_{1560}} f_e. \qquad (6)$$

Here $V_{780}$ is the estimated height of the peak in the amplified detector output voltage using the 780 nm filters (Fig. 6), which was proportional to the intensity of the elastic scattering signal. $V_{1560}$ was taken to be the uncertainty (one standard deviation) in the Gaussian fit to the "peak" in the 1560 scattering data (Fig. 7), which was consistent with zero. $A_{780}$ is the attenuation of the 780 signal by filter $F_5$, while $A_{1560} = 1$ since there was no attenuation of the 1560 signal. $L_{780}$ and $L_{1560}$ are the estimated loss factors (attenuation) due to the various optical elements, including the narrow-band filters, long-pass filters, and lens coatings, which were based on the manufacturer's test data combined with measurements made in the lab. $g_{780}$ and $g_{1560}$ are the external amplifier gains applied to the two signals, while $\eta_{780}$ and $\eta_{1560}$ are the detector quantum efficiencies at the two different wavelengths. Finally, $f_e$ represents the additional uncertainty in the ratio of the two scattering rates due to the uncertainties in the experimental parameters, which are summarized in Table 1.

TABLE I. Experimental parameters.

| parameter | 780 nm | 1560 nm |
|---|---|---|
| detector output (V) | 1.18±1% | 0.0012±15% |
| attenuation factor | 620±5% | 1 |
| loss factor | 1.60±4% | 2.05±1.4% |
| quantum efficiency | 0.15±5% | 0.75±2% |
| external gain | 10.0±1% | 100.0±1% |
| error factor $f_e$ | 1.17 | 1.17 |

Inserting the experimental parameters of Table I into Eq. (6) gives an upper bound on the scattering ratio of

$$R \leq 1.48 \times 10^{-7} \quad (7)$$

at the 99% confidence level. The generalized theory of Ref. [11] predicts a ratio of $R = 4\gamma^2$, provided that the mass of the oscillaton is negligibly small. Combining this with Eq. (7) gives an upper bound on the mixing angle $\gamma$ given by

$$|\gamma| \leq 1.93 \times 10^{-4} \text{ rad} \quad (8)$$

at the 99% confidence level.

## V. SUMMARY AND CONCLUSIONS

A generalization of quantum optics and quantum electrodynamics was recently suggested [11], in which the usual wave function $\psi_j(x_j)$ for a single mode $j$ of the electromagnetic field is further quantized to produce a field operator $\hat{\psi}_j(x_j)$. The operator $\hat{\psi}_j^\dagger(x_j)$ creates an additional hypothetical particle (oscillaton) in the harmonic oscillator corresponding to mode $j$ of the field. The generalized theory includes an unknown mixing angle $\gamma$ that couples the oscillaton creation and annihilation operators, as suggested by a symmetry breaking mechanism. The theory is equivalent to conventional quantum optics and quantum electrodynamics for $\gamma = 0$, but it predicts a new form of inelastic photon scattering for $\gamma \neq 0$.

Although there was no direct evidence for the existence of the proposed particles (oscillatons), there were several motivating factors for the theory:

- The third quantization technique is required for a complete description of quantum optics in the Heisenberg picture [11].
- The oscillatons may be candidates for the dark matter inferred from astronomical observations [23] if they have a large mass.
- The theory may be relevant to the discrepancies observed in recent QED experiments, such as the fine structure of positronium [15,16] and the magnetic moment of the muon [17,18].

In order to investigate the possible existence of these particles, an optical scattering experiment was performed in which the ratio of the predicted inelastic scattering rate divided by the usual elastic scattering rate was measured at an incident wavelength of 780 nm. No evidence for inelastic photon scattering of that kind was found, and the experiment sets an upper bound on the mixing angle of $|\gamma| \leq 1.93 \times 10^{-4}$ rad at the 99% confidence level, provided that the mass $m$ of the oscillaton is negligibly small.

If $m \neq 0$, then energy conservation requires that $\omega' = \omega/2 - mc^2/\hbar$, and no inelastic scattering of that kind is possible unless $\hbar\omega > 2mc^2$. As a result, high-energy photons from synchrotron radiation or cosmic ray showers may be required in order to observe the predicted inelastic scattering if $m$ is very large. It may be worth noting that the theoretical calculations of Ref. [11] used the nonrelativistic theory of a harmonic oscillator, which would only be valid in the limit of large $m$ and therefore low velocities.

It has been suggested that a null-result experiment of this kind does not merit publication. But Richard Feynman once said that "Anyone who performs an experiment has an obligation to publish the results [24]." That applies to null-result experiments as well, and there are examples of null-result experiments that played a major role in the development of physics, such as the Michelson-Morley experiment. The null result from this experiment suggests that further high-energy experiments may be required to test the theory.

In summary, the third quantization of the electromagnetic field is an interesting generalization of quantum electrodynamics that goes beyond the standard model. The results of the optical scattering experiment reported here set a relatively tight bound on the value of the mixing angle $\gamma$, provided that the mass of the oscillaton is negligibly small. High-energy experiments would be required to test the predictions of the theory if the mass of the oscillaton is very large.

## ACKNOWLEDGEMENT

This work was supported in part by the National Science Foundation under grant number PHY-1802472.